\documentclass[aip,apl,amsmath,amssymb,11pt,a4paper,longbibliography,reprint,notitlepage]{revtex4-1} 

\usepackage{graphicx,graphics,epsfig}
\usepackage[colorlinks=true,allcolors=blue]{hyperref}
\usepackage[utf8]{inputenc}
\usepackage[english]{babel}
\usepackage{color}
\usepackage{SIunits}
\usepackage[normalem]{ulem} 

\newcommand{\SM}{appendix}

\newcommand{\fref}[1]{Fig.~\ref{#1}}

\begin{document}

\title{Suspended photonic crystal membranes in AlGaAs heterostructures for integrated multi-element optomechanics}

\author{Sushanth Kini Manjeshwar}
\affiliation{Department of Microtechnology and Nanoscience (MC2), Chalmers University of Technology, SE-412 96 Gothenburg, Sweden}

\author{Karim Elkhouly}
\altaffiliation{Present address: IMEC, Kapeldreef 75, 3001 Leuven, Belgium}
\affiliation{Department of Microtechnology and Nanoscience (MC2), Chalmers University of Technology, SE-412 96 Gothenburg, Sweden}

\author{Jamie M.~Fitzgerald}
\affiliation{Department of Physics, Chalmers University of Technology, SE-412 96 Gothenburg, Sweden}

\author{Martin Ekman}
\affiliation{Department of Physics, Chalmers University of Technology, SE-412 96 Gothenburg, Sweden}

\author{Yanchao Zhang}
\affiliation{Zhejiang Supermat Sen-Ray Technology, No.~28, Middle Yunshan Road, 315400 Ningbo, China}

\author{Fan Zhang}
\affiliation{Zhejiang Supermat Sen-Ray Technology, No.~28, Middle Yunshan Road, 315400 Ningbo, China}

\author{Shu Min Wang}
\affiliation{Department of Microtechnology and Nanoscience (MC2), Chalmers University of Technology, SE-412 96 Gothenburg, Sweden}

\author{Philippe Tassin}
\affiliation{Department of Physics, Chalmers University of Technology, SE-412 96 Gothenburg, Sweden}

\author{Witlef Wieczorek}
\email{witlef.wieczorek@chalmers.se} 
\affiliation{Department of Microtechnology and Nanoscience (MC2), Chalmers University of Technology, SE-412 96 Gothenburg, Sweden}

\date{\today}

\begin{abstract}
    We present high-reflectivity mechanical resonators fabricated from AlGaAs heterostructures for use in free-space optical cavities operating in the telecom wavelength regime. The mechanical resonators are fabricated in slabs of GaAs and patterned with a photonic crystal to increase their out-of-plane reflectivity. Characterization of the mechanical modes reveals residual tensile stress in the GaAs device layer. This stress results in higher mechanical frequencies than in unstressed GaAs and can be used for strain engineering of mechanical dissipation. Simultaneously, we find that the finite waist of the incident optical beam leads to a dip in the reflectance spectrum. This feature originates from coupling to a guided resonance of the photonic crystal, an effect that must be taken into account when designing slabs of finite size. The single- and sub-$\upmu$m-spaced double-layer  {slabs} demonstrated here can be directly fabricated on top of a distributed Bragg reflector mirror in the same material platform. Such a platform opens a  route for realizing integrated multi-element cavity optomechanical devices and optomechanical microcavities on chip.
\end{abstract}


\maketitle 

Cavity optomechanical devices explore the interaction between light and mechanical resonators in a cavity \cite{Aspelmeyer2014} and rely on strongly coupled, high-quality optical and mechanical resonators. When several independent mechanical resonators are coupled to a single cavity field, one is in the realm of multi-element optomechanics \cite{xuereb_strong_2012,Xuereb2013}, which has been proposed as a route to reach the elusive single-photon strong optomechanical coupling regime \cite{Rabl2011,nunnenkamp_single-photon_2011}. Recent experiments along these lines \cite{Nair2017,piergentili_two-membrane_2018,Gartner2018,wei_controllable_2019} have used SiN membranes placed in free-space optical cavities, but require precise alignment of their tilt angle and position and, additionally, a uniformity of the mechanical and optical properties of individual membranes.

Using III-V heterostructures such as AlGaAs would allow for the realization of a multi-element cavity optomechanical system in a fully integrated approach \cite{vanner_pulsed_2011,cole_tensile-strained_2014}. A heterostructure can integrate one of the cavity mirrors via a distributed Bragg reflector together with an array of near-uniform mechanical resonators on a single wafer \cite{roh_strong_2010,vanner_pulsed_2011,cole_tensile-strained_2014}, and can even be combined with micro-mirrors on an independent chip \cite{wachter_silicon_2019}. In particular, the III-V materials system has already been used to realize (opto)mechanical systems in, e.g., (Al)GaAs  \cite{yamaguchi_improved_2008,watanabe_feedback_2010,ding_high_2010,Cole2010,cole_phonon-tunnelling_2011,okamoto_vibration_2011,Liu2011,usami_optical_2012,midolo_design_2014,yamaguchi_gaas-based_2017,hamoumi_microscopic_2018,cotrufo_nanomechanical_2018} or In(Ga)P \cite{antoni_deformable_2011,cole_tensile-strained_2014,makles_2d_2015,Buckle2018}. These crystalline materials have been shown to be of high optical \cite{cole_tensile-strained_2014,cole_high-performance_2016} and mechanical quality \cite{Liu2011,cole_tensile-strained_2014,hamoumi_microscopic_2018} as required for cavity optomechanics. Further device functionalization based on the piezoelectricity of III-V materials or by embedding quantum emitters can lead to versatile nano-electro-optomechanical systems \cite{midolo_nano-opto-electro-mechanical_2018}. 

In this Letter, we demonstrate the fabrication of single- and double-layer  {membranes} in AlGaAs heterostructures \cite{Manfra2014} {and present a comprehensive characterization of single-layer high-reflectivity mechanical resonators}. The mechanical resonators are fabricated in \unit{100}{\nano\meter}-thin GaAs membranes, which are grown on top of sacrificial AlGaAs layers. This allows us to fabricate double-layer  {membranes} with sub-$\upmu$m spacing, which is crucial for reaching high coupling strengths in multi-element optomechanics \cite{xuereb_strong_2012,Xuereb2013}. We engineer mechanical resonators of free-free-type geometry \cite{Cole2010,cole_tensile-strained_2014} and characterize {their} mechanical properties. We demonstrate control over their out-of-plane optical reflectivity in the telecom wavelength regime by patterning a photonic crystal (PhC) into the GaAs membranes \cite{fan_analysis_2002}, as has been demonstrated in optomechanics \cite{bui_high-reflectivity_2012,makles_2d_2015,norte_mechanical_2016,bernard_precision_2016,Chen2017,moura_centimeter-scale_2018} and optical communication technologies \cite{Stomeo2010,zhou_progress_2014}. Our devices constitute a significant step towards the realization of an array of near-uniform mechanical resonators integrated in a free-space, fully chip-based cavity optomechanical device.

\begin{figure}[b!thp]
\centering\includegraphics[width=0.8\linewidth]{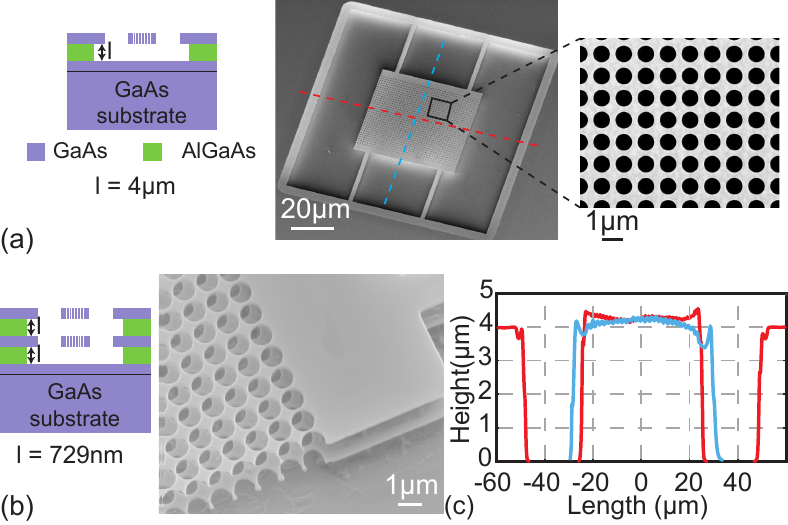}
\caption{Free-free-type mechanical resonators fabricated in AlGaAs heterostructures. (a)~Schematic of the AlGaAs heterostructure and scanning electron microscope (SEM) images of a single-layer free-free-type mechanical resonator structured as a photonic crystal (PhC) membrane. (b)~Schematic and SEM image of a double-layer PhC device. (c)~Height profile of the device from (a).}
\label{fig:SEMofDevices}
\end{figure}

The mechanically-compliant PhC slabs are fabricated in an AlGaAs heterostructure that is epitaxially grown on a GaAs substrate using molecular-beam epitaxy. We fabricated devices from two different wafers. The heterostructure of the first wafer is used for single-layer mechanical resonators [\fref{fig:SEMofDevices}(a)]. It consists of a {\unit{100}{\nano\meter}-thick} GaAs device layer grown on top of a  \unit{4}{\micro\meter}-thick Al$_{0.65}$Ga$_{0.35}$As sacrificial layer. The AlGaAs layer exhibited a large peak-to-peak surface height variation of \unit{15}{\nano\meter} that is partially smoothened by the top GaAs layer to \unit{10}{\nano\meter} {yielding an average GaAs layer thickness of 89\,nm inferred from transmission electron microscopy (TEM) and ellipsometry measurements}. 

The second wafer is designed for fabricating sub-$\upmu$m spaced, double-layer GaAs mechanical devices [\fref{fig:SEMofDevices}(b)], each of \unit{100}{\nano\meter} thickness on top of a \unit{729}{\nano\meter} Al$_{0.625}$Ga$_{0.375}$As sacrificial layer, which defines the spacing between the two {GaAs slabs}. These AlGaAs layers were grown with growth interruption, \cite{sweet_gaas_2010} yielding a surface height variation and roughness smaller than \unit{1}{\nano\meter} and \unit{0.2}{\nano\meter}, respectively. We used standard AlGaAs heterostructure\cite{Stomeo2010,midolo_soft-mask_2015} microfabrication techniques to define the patterned mechanical resonators and their release (see the \SM{} for {details}).

\begin{figure}[b!thp]
\centering\includegraphics[width=0.45\textwidth]{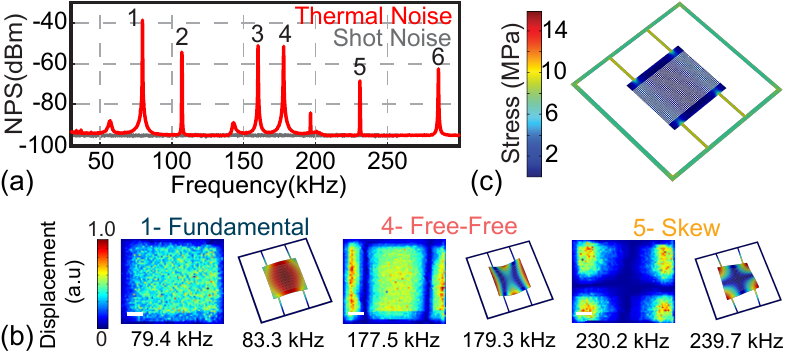}
\caption{Characterization of mechanical modes of a free-free-type PhC slab. (a)~Noise power spectrum (NPS) of the thermally driven mechanical motion (red) with mechanical modes labeled 1 to 6. (b)~Mechanical-mode tomography of the same device along with FEM-simulated mode shapes and their frequencies. Scale bar: \unit{10}{\micro\meter}. Note that the device boundary inferred from mode tomography is largely determined by the rectangular \unit{50\times38}{\micro\meter\squared} PhC area that reflects more light than the non-patterned part and, thus, leads to an apparent deviation from the square shape of the slab. (c)~FEM-simulated von Mises stress distribution of the device.}
\label{fig:NPS_50x50_PhC}
\end{figure}

{The} mechanical resonators are engineered with a free-free-type geometry \cite{Cole2010}, where the suspended slab is of rectangular shape and held by four tethers at the nodes of the free-free oscillation mode \cite{cole_phonon-tunnelling_2011} [see \fref{fig:NPS_50x50_PhC}(b)]. Firstly, we characterized the mechanical properties of the slabs, focusing on the mode shapes and corresponding eigenfrequencies and quality factors. To this end, we detected the out-of-plane displacement of the slab via optical homodyne interferometry at room temperature in a high vacuum ($\sim\unit{5\times10^{-5}}{\milli\bbar}$)---for details of the setup {see} the \SM{}. \fref{fig:NPS_50x50_PhC}(a) shows a typical displacement noise power spectrum of a PhC slab with a size of $\unit{50\times50\times0.1}{\micro\meter\cubed}$. The fundamental mode lies at \unit{80}{\kilo\hertz} and the free-free mode at \unit{178}{\kilo\hertz}. Mechanical mode tomography \cite{barg_measuring_2016} enabled us to compare the measured mode shape to finite element modeling (FEM) simulations\cite{comsol}, see \fref{fig:NPS_50x50_PhC}(b). We find good agreement between experimental and FEM data when accounting for a tensile stress of \unit{10}{\mega\pascal} in the GaAs layer.

We attribute the residual tensile stress to a mismatch between the lattice constants of the AlGaAs sacrificial and the GaAs device epilayers. The AlGaAs grown on the GaAs substrate relaxes to its native lattice constant if the layer thickness exceeds the critical thickness of \unit{0.33}{\micro\meter} or \unit{30}{\micro\meter}, according to Ref.~\cite{Matthews1970} or Ref.~\cite{People1985}, respectively (see also the \SM{}). These predictions differ by two orders of magnitude such that the AlGaAs layer can be in a state between fully relaxed and fully strained, depending on the model used \cite{Matthews1970,People1985}. The GaAs device layer is thinner than its critical thickness and, thus, adapts to the lattice constant of the AlGaAs layer in any case. Then, the GaAs device layer can exhibit a tensile stress of between
0 and \unit{77.5}{}{\mega\pascal}.

\begin{figure}[b!thp]
    \centering\includegraphics[width=0.45\textwidth]{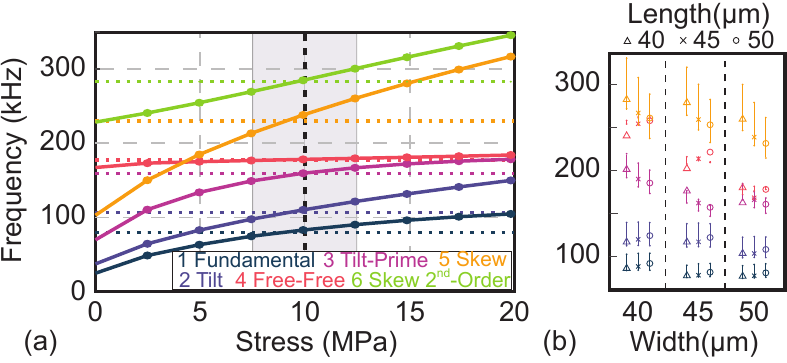}
    \caption{(a)~FEM simulation results for mechanical frequencies of a \unit{50\times50}{\micro\meter\squared} PhC patterned device with varying tensile stress in a \unit{105}{\nano\meter}-thick GaAs layer (lines are guides to the eye). The dotted horizontal lines show the measured frequencies from \fref{fig:NPS_50x50_PhC}(a). (b)~Measured frequencies for patterned devices of different dimensions. The bars around each data point denote the FEM-simulated frequency range corresponding to the stress range marked by the shaded region in (a).}
    \label{fig:PhCStrainSweep}
\end{figure}

In \fref{fig:PhCStrainSweep}(a) we examine the effect of tensile stress in the GaAs device layer on the eigenfrequencies of the suspended slab. We observe that the frequencies increase with stress and find a match between data and FEM for a stress around \unit{10}{\mega\pascal}. Upon removal of the sacrificial AlGaAs layer, an anisotropic stress distribution develops in the suspended GaAs slab, as shown in \fref{fig:NPS_50x50_PhC}(c). We also observe buckling of the slabs \cite{Shevyrin2012} and a static deformation \cite{Barg2018}. {For example, } \fref{fig:SEMofDevices}(c) {shows a slab with 280\,nm buckling and a static deformation of 400\,nm along the direction indicated by the red line}. We conclude that the GaAs layer exhibits residual tensile stress induced by the underlying AlGaAs layer, as was observed in other GaAs on AlGaAs resonators \cite{Liu2011,Shevyrin2012,Barg2018}. 

Spatial variations of the resonator geometry \cite{Shevyrin2012} or defect-driven material anisotropy \cite{Buckle2018} also influence the mechanical properties. While analysis of the latter is beyond the scope of our work, the former can be caused by growth-related thickness variation or microfabrication-induced changes. Geometry variations influence the mode-dependent oscillating mass of the resonator and, thus, its eigenfrequencies. We account for the geometry of the devices in FEM with the simplifying assumption of a constant GaAs layer thickness. \fref{fig:PhCStrainSweep}(b) shows frequencies for devices of various slab length and width. For a layer thickness of \unit{105}{\nano\meter} {assumed in the FEM}, we find good agreement between measured and FEM-simulated frequencies (see the \SM{} for detailed simulation results). {We attribute the thickness difference of \unit{15}{\nano\meter} between FEM and the TEM\slash ellipsometry measurements to the assumption in the FEM that the GaAs layer exhibits a constant thickness, which simplifies modeling, but neglects spatial mass variations of the slab.} Overall, residual tensile stress in the GaAs layer and the simplifying assumption of its constant thickness yields a reasonable explanation for the observed mechanical frequencies of the suspended PhC slabs.

The mechanical quality factor, $Q$, is an important figure of merit for (opto)mechanical devices. We find that devices fabricated from the first wafer have quality factors just below $10^{5}$ and similar devices from the second wafer reach $3 \times 10^{5}$ (see the \SM{} for all data), which is about a factor of 10 larger (4 smaller) than Refs.~\cite{cole_phonon-tunnelling_2011,Barg2018} (Ref.~\cite{Liu2011}). We do not observe any systematic discrepancy between the $Q$ of patterned and unpatterned devices. We expect an increase in $Q$ by at least an order of magnitude when using samples with smoother surfaces \cite{cole_high-performance_2016}, operating at lower temperatures \cite{Barg2018,hamoumi_microscopic_2018}, and using strain engineering \cite{norte_mechanical_2016,tsaturyan_ultracoherent_2017,ghadimi_elastic_2018}.

\begin{figure}[t!bhp] 
\includegraphics{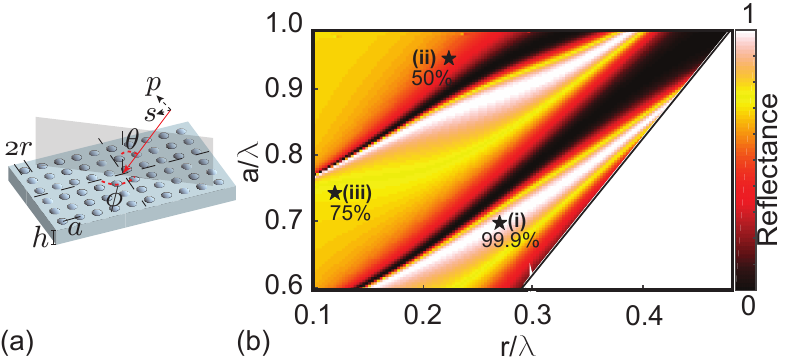}
\caption{(a) Schematic of a PhC slab of thickness $h$ with a square PhC of lattice constant $a$ and hole radius $r$. A plane wave is incident at polar angle $\theta$ and azimuthal angle $\phi$ with polarization components $s$ and $p$. (b) Reflectance map for a $\lambda = \unit{1550}{\nano\meter}$ plane wave at normal incidence on a $h=\unit{100}{\nano\meter}$ GaAs PhC slab for varying $r$ and $a$. The stars mark PhC patterns with (i)~$a = \unit{1081}{\nano\meter}$, $r = \unit{418}{\nano\meter}$ ($R>99$\%), (ii)~$a = \unit{1452.8}{\nano\meter}$, $r = \unit{318.8}{\nano\meter}$ ($R=50$\%) , and (iii)~$a = \unit{1162.8}{\nano\meter}$, $r = \unit{159.18}{\nano\meter}$ ($R=75$\%).}
\label{fig:PhC_Overview}
\end{figure}

The unpatterned membrane has an out-of-plane optical reflectance of 69\% at a free-space wavelength of \unit{1550}{\nano\meter}, which is too low for reaching single-photon strong coupling in a multi-element optomechanical device \cite{xuereb_strong_2012,Xuereb2013}. By patterning the membranes as a PhC with air holes arranged in a square lattice, \cite{fan_analysis_2002,bui_high-reflectivity_2012,makles_2d_2015,norte_mechanical_2016,Chen2017} we can engineer a reflectance between 0\,\% and 100\,\% (\fref{fig:PhC_Overview}), which we calculated using rigorous coupled wave analysis, e.g., via the S4 package \cite{Liu20122233}. To demonstrate this capability, we fabricated devices aiming at a reflectance of (i) $99$\,\%, (ii) $75$\,\% and (iii) $50$\,\% at \unit{1550}{\nano\meter}.

\begin{figure}[t!bhp]
\centering\includegraphics[width=0.8\linewidth]{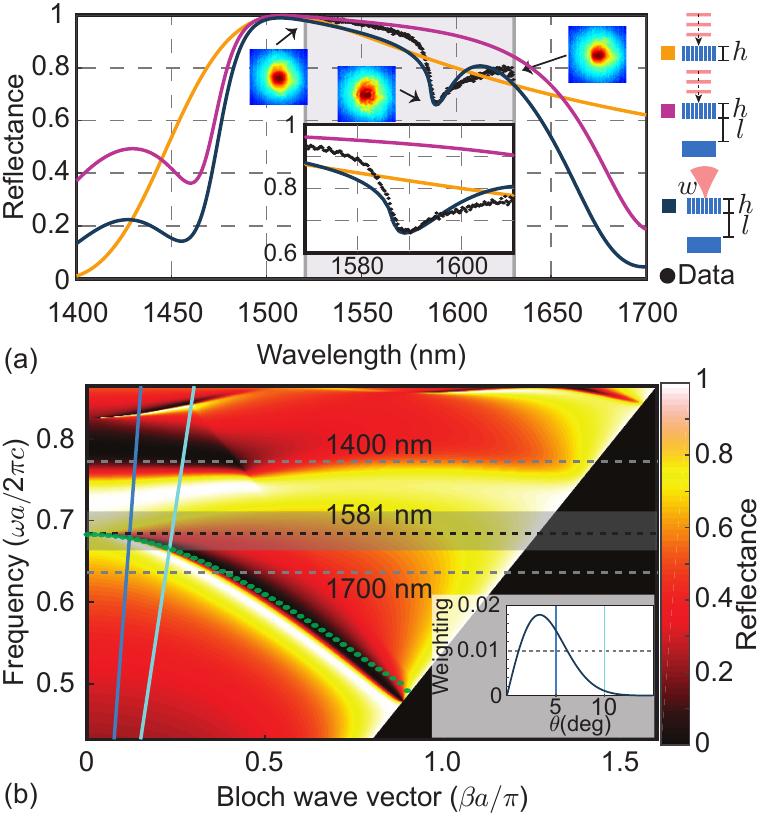}
\caption{(a)~Reflectance spectra of a suspended GaAs PhC membrane of thickness $h = \unit{87.5}{\nano\meter}$ and air-gap $l = \unit{4.3}{\micro\meter}$. The data (black) are compared to simulated spectra for a plane wave/Gaussian beam of waist $\unit{4.2}{\micro\meter}$ incident on the PhC slab (orange)/(-) or on the slab on top of a GaAs substrate (purple)/(blue). The gray region marks the measurement range. The insets show transverse mode patterns measured in reflection. (b)~Reflectance map of a PhC membrane for an incident plane wave of wave vector $\beta=\frac{2\pi}{\lambda}\sin(\theta)$ and frequency $\omega$. The green dots show the dispersion of the guided resonance for an $s$-polarized wave. The inset shows the weighting factor of plane waves used for representing a Gaussian beam of waist $\unit{4.2}{\micro\meter}$. The blue lines mark the same angles of incidence in the inset and main panel.}
\label{fig:99pc_SimpleSpectra}
\end{figure}

We focus on device (i) in \fref{fig:99pc_SimpleSpectra} and discuss devices (ii) and (iii) in the \SM{} along with a description of the optical setup used to measure reflectance \cite{rivoire_gallium_2008}. In \fref{fig:99pc_SimpleSpectra}(a), we observe a maximum of the reflectance around \unit{1510}{\nano\meter}, away from the designed maximum at \unit{1550}{\nano\meter}. {We  reproduce this shift for a slab with a thickness of \unit{87.5}{\nano\meter} in the PhC simulation (instead of the assumed \unit{100}{\nano\meter}), which closely matches the thickness of 89\,nm inferred from the TEM and ellipsometry measurements.}

Notably, the reflectance spectrum in \fref{fig:99pc_SimpleSpectra}(a) shows a pronounced dip at 
\unit{1581}{\nano\meter}. This dip can only be reproduced when taking into account the finite waist of the incident beam. To this end, we model the incident Gaussian beam as a weighted sum of plane waves incident at polar angle $\theta$ and azimuthal angle $\phi$  [\fref{fig:PhC_Overview}(a) and inset \fref{fig:99pc_SimpleSpectra}(b)], see Refs.~\cite{singh_chadha_polarization-_2013,moura_centimeter-scale_2018} and the \SM{}. The dip results from coupling of plane waves at oblique incidence to a guided resonance of the PhC. This can be best illustrated with the reflectance map of the PhC slab shown in \fref{fig:99pc_SimpleSpectra}(b). The dispersion relation of the guided resonance at \unit{1581}{\nano\meter} at wave vector $\beta=0$ shows a decrease in frequency with an increase in $\beta$. Hence, the guided resonance appears at longer wavelengths for light impinging under oblique incidence. As a Gaussian beam is formed by the weighted sum of many plane waves, a reflectance dip of finite spectral width is formed.

\begin{figure}[t!bhp]
\centering\includegraphics[width=0.8\linewidth]{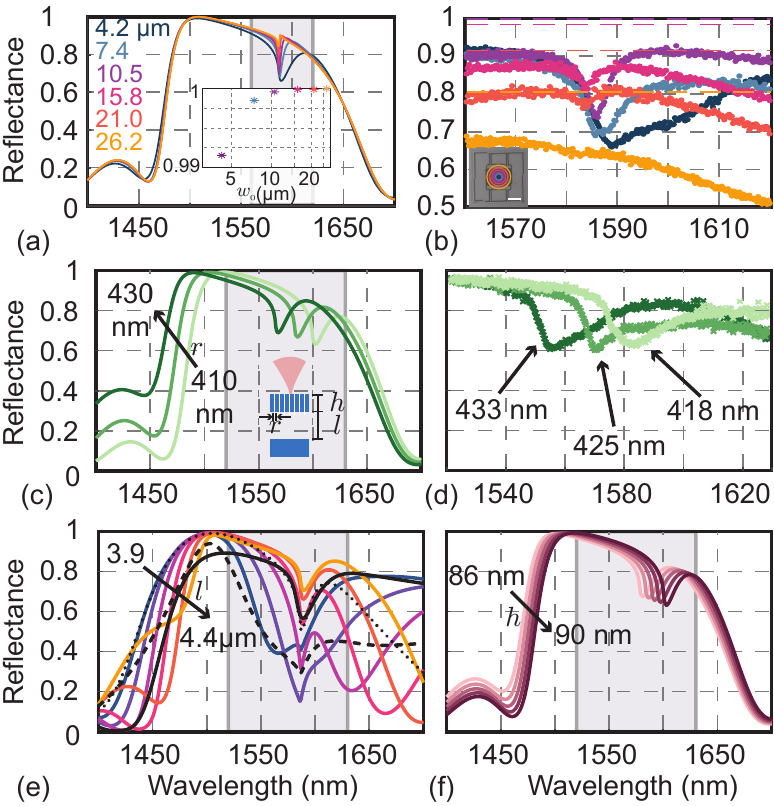}
\caption{(a)~Reflectance spectra of a Gaussian beam of varying waist size incident on an infinite PhC slab on top of a substrate. The inset shows the maximal achievable reflectance for a given waist. (b)~Measured  spectra for different waist sizes. The horizontal dashed lines represent the expected clipping loss, see inset. Scale bar: 25$\,\mu$m. (c)~Calculated and (d)~measured reflectance spectra when varying the PhC hole radius $r$. Calculated reflectance spectrum when varying (e)~the air-gap spacing $l$ {(with black solid, dashed and dotted lines represent $l=0.5$, 0.75 and 1\,$\mu$m, respectively)} and (f)~slab thickness $h$.}
\label{fig:Reflectivity_all_99}
\end{figure}

In \fref{fig:Reflectivity_all_99}(a) we examine the effect of varying waist on reflectance. For larger waists, the dip in the spectrum narrows. The reason for this behavior is that larger waists are represented by plane waves with weighting factors that favor less oblique contributions and, thus, less dispersion of the guided resonance is collected. Furthermore, a larger waist reaches a larger reflectance \cite{moura_centimeter-scale_2018}, as seen in the inset of \fref{fig:Reflectivity_all_99}(a). In our measurements in \fref{fig:Reflectivity_all_99}(b), we observe that the dip width indeed decreases with increasing waist. However, in contrast to our prediction, we observe an overall drop in reflectance with larger waists. We attribute this drop to clipping loss due to the finite size of the slab and to diffraction loss of the guided resonance at the boundaries of the slab \cite{grepstad_finite-size_2013}.

Finally, we study the dependence of reflectance on parameters of the PhC device. \fref{fig:Reflectivity_all_99}(c) shows that the dip shifts to shorter wavelengths upon increasing the radius of the PhC holes. This shift is expected as the patterning determines the {PhC} mode structure. We illustrate this behavior with three devices in \fref{fig:Reflectivity_all_99}(d). \fref{fig:Reflectivity_all_99}(e) shows the dependence on the air-gap $l$. We observe that the position of the dip remains constant, as expected since the mode structure of the PhC {slab} is not influenced by {$l$}. However, the reflectance at {the dip} depends strongly on {$l$}. This is the result of a spectral shift of the Fabry-P\'erot resonance formed by the slab and substrate through the dip. 
\fref{fig:Reflectivity_all_99}(f) shows the dependence on slab thickness $h$. We observe that the dip shifts to longer wavelengths with increasing $h$, with a strong shift of \unit{4.5}{\nano\meter} in wavelength per nm change in thickness. Hence, a precise knowledge of the slab thickness is required to engineer the position of the dip accurately. 

To conclude, we have demonstrated the engineering of suspended PhC slabs in GaAs with mechanical resonance frequencies above \unit{50}{\kilo\hertz}, quality factors as high as $3\times 10^5$ at room temperature and a maximal $Q\times f$ product of \unit{10^{11}}{\hertz}, and a controllable out-of-plane reflectance at telecom wavelengths\cite{fan_analysis_2002,norte_mechanical_2016,bernard_precision_2016}. The GaAs device layer exhibited residual tensile stress, which can be favorably used for strain engineering to reduce mechanical dissipation as demonstrated, e.g., with SiN \cite{verbridge_high_2006,norte_mechanical_2016,Reinhardt2016,tsaturyan_ultracoherent_2017,ghadimi_elastic_2018} or III-V-based resonators \cite{watanabe_feedback_2010,onomitsu_ultrahigh-q_2013,cole_tensile-strained_2014,Buckle2018}. A dip in the reflectance spectrum \cite{singh_chadha_polarization-_2013,bernard_precision_2016,moura_centimeter-scale_2018} originating from coupling to a guided resonance in the PhC was observed. Hence, PhC devices of finite size must be carefully engineered to have this dip outside of a desired high-reflectivity region.

The mechanical resonator slabs in GaAs presented in this Letter can be engineered into arrays of high-reflectivity mechanical resonators of precise, epitaxially defined thickness and spacing using AlGaAs heterostructures integrated on top of a distributed Bragg reflector \cite{vanner_pulsed_2011,Barg2018}. Such an integrated system presents {exciting} perspectives for realizing free-space and fully chip-based multi-element cavity optomechanical systems \cite{xuereb_strong_2012,Nair2017,piergentili_two-membrane_2018,Gartner2018,wei_controllable_2019}, optomechanical microcavities \cite{vanner_pulsed_2011}, or frequency-dependent mirrors in optical cavities  \cite{naesby_microcavities_2018,cernotik_cavity_2019}.

{See \SM{} for detailed descriptions of device fabrication, experimental setups, FEM simulations, modeling of PhC reflectance, and further data.}

We acknowledge fruitful discussions with Claus G\"artner, Garrett D.~Cole and Martí Gutierrez Latorre. The work was supported by Chalmers’ Excellence Initiative Nano, the Knut and Alice Wallenberg Foundation, the Swedish Research Council, and the QuantERA project C'MON-QSENS!. Samples were fabricated in the Myfab Nanofabrication Laboratory at Chalmers and analyzed in the Chalmers Materials Analysis Laboratory. Simulations were performed on resources provided by the Swedish National Infrastructure for Computing at C3SE.

The data that support {our} findings are openly available on zenodo.org, DOI: \href{https://doi.org/10.5281/zenodo.3783499}{10.5281/zenodo.3783499}.

\appendix

\section{Device concept}

\subsection{Device fabrication}
\begin{figure}[t!bp]
\centering\includegraphics{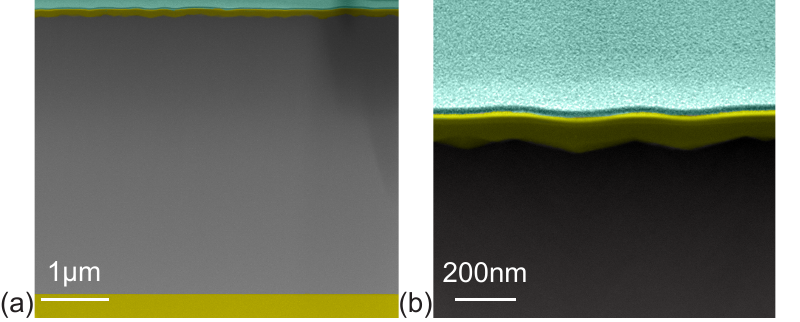}
\caption{False-coloured transmission electron microscopy image of the MBE-grown AlGaAs heterostructure. The top layer (green) is palladium deposition for ion beam milling, followed by the GaAs layer (yellow) and the AlGaAs layer (gray) on top of the GaAs substrate (yellow).}
\label{fig:TEMofMBE}
\end{figure}

The mechanically compliant PhC slabs were fabricated in an AlGaAs heterostructure epitaxially grown using molecular-beam epitaxy (MBE) on a $\langle 100 \rangle$ oriented GaAs substrate, where the first grown layer is a \unit{100}{\nano\meter} GaAs buffer layer. The AlGaAs layer of the first wafer was not grown with growth interruption. As a result, the top surface layer had a large surface height variation, as seen in \fref{fig:TEMofMBE}. {From \fref{fig:TEMofMBE} we infer an average thickness of the GaAs layer of \unit{89.6}{\nano\meter} and a maximal and minimal thickness of \unit{105.6}{\nano\meter} and \unit{69.4}{\nano\meter}, respectively. Complementary, we examined the as-grown structure with an ellipsometer and find a similar layer thickness of $(89.1\pm0.5)$\,nm. Note that the fabricated and released devices show buckling and static deformation. For example, the device shown in Fig.~1(c) of the main text shows an average buckling above the substrate of \unit{280}{\nano\meter} perpendicular to the direction of the tethers and an average buckling above the substrate of \unit{85}{\nano\meter} along the direction of the tethers. The static deformation has the approximate shape of a parabola with a peak-to-peak deflection of about \unit{400}{\nano\meter} perpendicular to the direction of the tethers and \unit{560}{\nano\meter} along the direction of the tethers.}

The AlGaAs layers of the second wafer were grown with a growth interruption \cite{sweet_gaas_2010} of \unit{60}{\second} after \unit{150}{\nano\meter} of deposited AlGaAs material, which lead to a marked improvement of height variation (less than \unit{1}{\nano\meter}) and root-mean-square surface roughness (less than \unit{0.2}{\nano\meter}).

We defined the geometry of the mechanical resonator and the PhC pattern using electron beam lithography with UV-60 resist. The pattern is transferred onto the device layer by inductively coupled plasma reactive ion etching (ICP-RIE) using SiCl$_{4}$/Ar chemistry \cite{Stomeo2010}. The sacrificial AlGaAs layers are removed by HF wet etching with an approximate etch rate of \unit{1}{\micro\meter\per\minute} followed by removal of etch remnants using KOH \cite{midolo_soft-mask_2015}. The openings of the chosen geometry of the mechanical resonator allow the etch-products of wet etching to be flushed out, which is crucial for fabricating multi-layer devices. Finally, the devices are dried using CO$_2$-based critical point drying to prevent stiction of the released slabs. We fabricated single-layer devices from the second wafer by stripping the top GaAs layer with SiCl$_{4}$/Ar ICP-RIE followed by HF wet etch to remove the top AlGaAs layer.

\subsection{Device scalability}

{Extending our approach to more than two membranes can be based on the concept of the presented AlGaAs heterostructure architecture. One would grow a structure of repeated sacrificial and device layers, very similar to growing a distributed Bragg reflector mirror (DBR). The overall growth would be limited by the total acceptable growth thickness in MBE that would still yield high quality epitaxial material. The longer the growth time in MBE, the more defects and contaminants are incorporated in the material and the lower is the device quality. A conservative estimate of this thickness yields about \unit{15}{\micro\meter}, see, e.g., Ref.~\cite{cole_high-performance_2016} for growth of DBR mirrors of about \unit{17}{\micro\meter} thickness grown with AlGaAs heterostructures. Assuming a device layer thickness of \unit{100}{\nano\meter} and a sacrificial layer thickness of \unit{750}{\nano\meter} this would yield an array of 17 membranes. If one were to integrate the mechanical array on top of a DBR, then this would reduce to an array of 8 membranes, when assuming a total DBR thickness of \unit{8}{\micro\meter}  optimized for a reflectivity at \unit{1550}{\nano\meter}.}

{A challenge, however, lies in the subsequent fabrication process. This process requires etching through all device layers via a dry etch process in order to define the lateral geometry of the devices. Subsequently, the wet etch process removes the AlGaAs material of all sacrificial layers. The wet etch process is helped by a clever choice of geometry of devices, which leaves enough space for the etch remnants to be removed from between the GaAs device layers. The dry etch process should guarantee vertical sidewalls during etching through all layers up to the last sacrificial layer. ICP etching of AlGaAs using Cl-based gases can etch about \unit{1}{\micro\meter}  deep with large aspect ratio \cite{zhang_dry_2017}, while reactive ion beam etching (RIBE) may offer etching depths up to \unit{1.7}{\micro\meter}  \cite{avary_deeply_2002}. If we assume a maximal etch depth of around \unit{2}{\micro\meter}, then we will be limited to the fabrication of 3 membranes.}

\begin{figure}[t!bhp]
\centering\includegraphics[width=.66\linewidth]{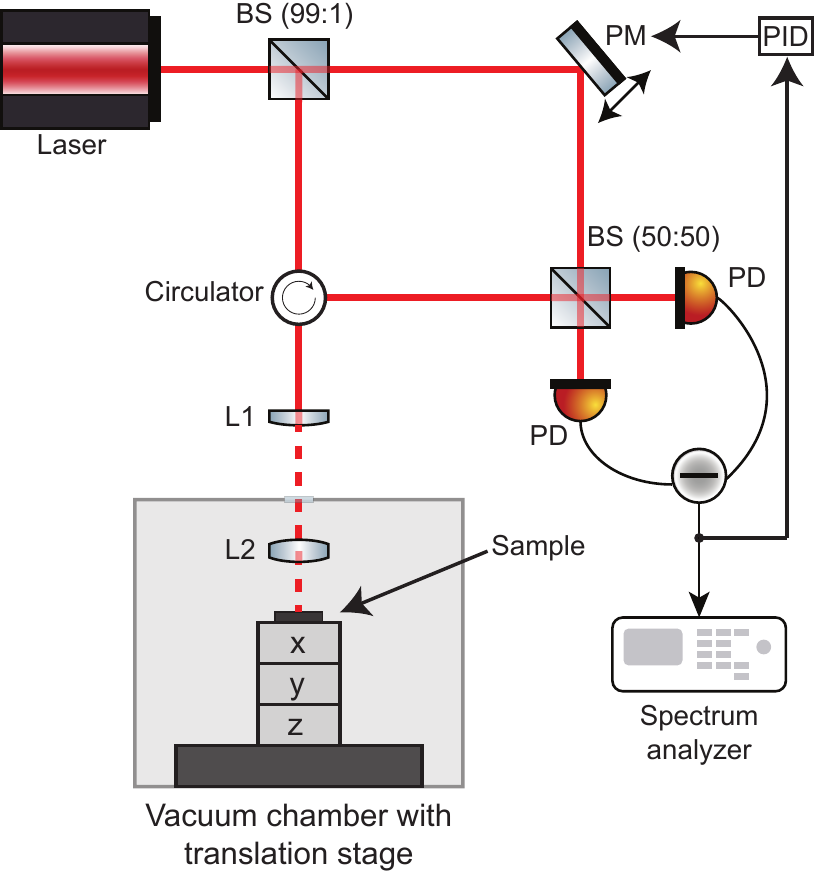}
\caption{Experimental setup for characterization of mechanical properties of the suspended PhC slabs. The solid lines indicate the fiber beam path and the dashed lines are the free-space beam path. BS: fiber-based beam splitter, PM: fiber-based phase modulator, PD: photodetector,  L1: collimator of focal length \unit{50}{\milli\meter}, L2: focal length \unit{20}{\milli\meter}.}
\label{fig:Mechanical_Setup}
\end{figure}

\subsection{Optical absorption}

{Optical absorption in the membrane material leads to undesired optical loss, which would, for example, reduce the Finesse of an optical cavity, where the presented devices would be placed in. We operate at a wavelength of \unit{1550}{\nano\meter}  and the GaAs membrane is about \unit{100}{\nano\meter} thick. We estimate that the optical absorption for this case should be below 5\,ppm. This estimate is taken from Ref.~\cite{michael_wavelength-_2007}, which measured absorption in GaAs optical resonators at telecom wavelengths. Absorption of a single layer device would then limit the cavity Finesse to be below $10^6$, which is much larger than an expected cavity Finesse when using high-reflectivity mirrors, which lies usually between values of $10^4$ to $10^5$.}

\section{Experimental setup}

The characterization of the mechanical properties of the suspended PhC slabs is carried out with an optical homodyne detection setup operating in the telecom wavelength regime, shown in \fref{fig:Mechanical_Setup}. A diode laser tunable between \unit{1520}{\nano\meter} and \unit{1630}{\nano\meter} is split into a signal and a local oscillator (LO) beam path using a fiber beam splitter. {The laser beam is focused onto the sample with a beam waist of $2.2\,\mu$m. } The signal light is reflected off the device, which is placed inside a vacuum chamber on an xyz translation stage. The displacements of the mechanical modes of the device imprint a phase shift on the reflected light beam. The reflected signal is then mixed with the local oscillator beam in a tunable fiber beam splitter, whose outputs are directed to a balanced photo receiver. The optical interferometer is locked on the phase quadrature by sending a feed back signal to a fiber-based phase modulator in the LO beam path. The electronic signal of the photo receiver is then analyzed with an electronic spectrum analyzer. {The mechanical mode tomography is performed by translating the sample in $1\,\mu$m steps along the x and y direction. The laser spot yields a spatial resolution of about $4.5\,\mu$m in diameter.}

\begin{figure}[t!bhp]
\centering\includegraphics[width=.66\linewidth]{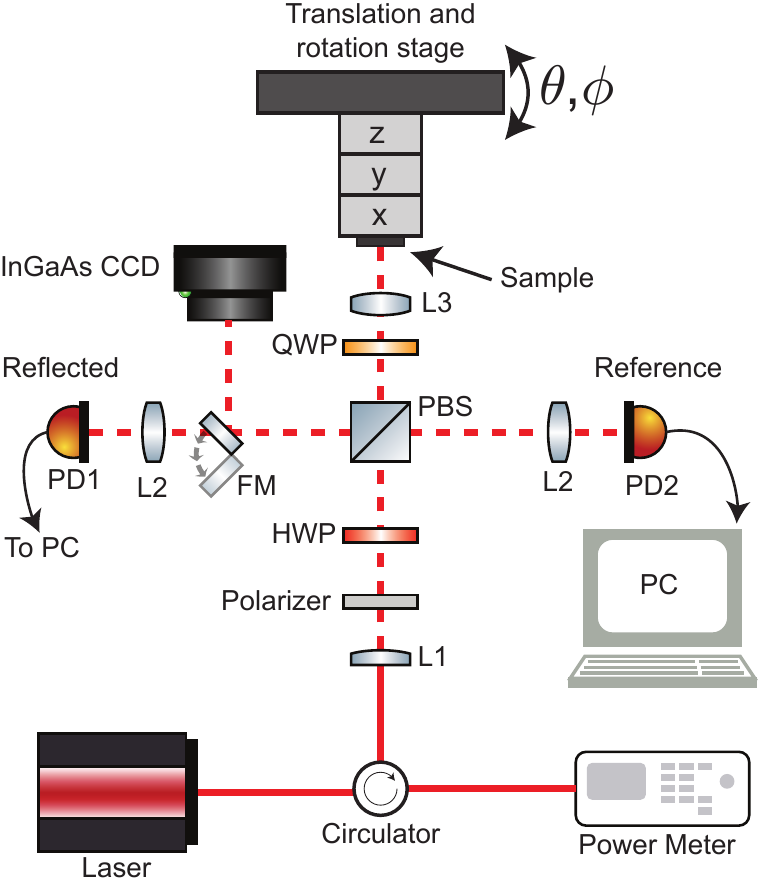}
\caption{Experimental setup for characterization of the optical reflectance of samples. The solid lines indicate the fiber beam path and the dashed lines are the free-space beam path. HWP: half-wave plate, QWP: quarter-wave plate, PBS: polarizing beam splitter, FM: flip mirror, PD: photodetector. L1: collimator with focal length \unit{25.2}{\milli\meter}, L2: focal length \unit{25.4}{\milli\meter}, L3: focal length \unit{20,35,50,75,100}{\milli\meter}  or \unit{125}{\milli\meter}.}
\label{fig:OpticalSetup}
\end{figure}

The setup for measuring the optical reflectance of the PhC slabs is shown in \fref{fig:OpticalSetup}. The laser light is first passed through a polarizer. A half-wave plate is used to adjust the ratio of the light reflected off a polarizing beam splitter and going to the reference arm and the light transmitted and going to the sample. A quarter-wave plate rotates the transmitted light to circular polarization. We use an aspheric lens to focus the light onto the PhC slab. By using aspheric lenses of different focal length, we can focus the beam to different beam waists on the sample. The light reflected off the sample collects a $\pi$-phase shift upon reflection and, after passing through the quarter-wave plate, is vertically polarized and, thus, reflected by the PBS into the detection arm. This type of reflectance measurement assumes that the reflection of light from the sample behaves the same for $s$- and $p$-polarized light. This is the case for our samples that use non-patterned devices or devices patterned with a PhC of $C_4$ symmetry.

The reflectance of the PhC slab, $R_{\mathrm{PhC}}$, is then given by

\begin{equation}
    R_\mathrm{PhC} = \left(\frac{I_{\mathrm{PD1}}^{\mathrm{PhC}}}{I_{\mathrm{PD2}}^{\mathrm{PhC}}}\right)\cdot\left[\left(\frac{I_{\mathrm{PD1}}^{\mathrm{Mirror}}}{I_{\mathrm{PD2}}^{\mathrm{Mirror}}}\right)^{\mathrm{-1}}\cdot R_{\mathrm{mirror}}\right]
    \label{eq:Reflectance_PhC}
\end{equation}

\begin{figure}[t!bhp]
\centering\includegraphics[width=.8\linewidth]{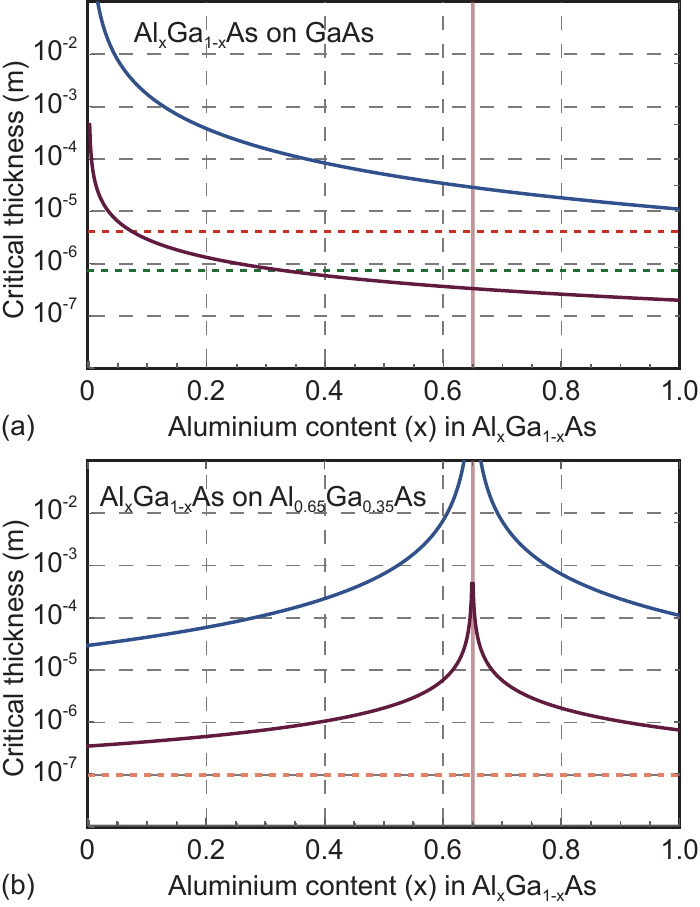}
\caption{Critical thickness calculated for (a) Al$_{x}$Ga$_{1-x}$As on GaAs and (b) Al$_{x}$Ga$_{1-x}$As on Al$_{0.65}$Ga$_{0.35}$As. The blue line is for the People and Bean model \cite{People1985}, while the brown line is for the Matthews model \cite{Matthews1970}. The horizontal lines in (a) represent the two thicknesses of AlGaAs used in this work, i.e., $4\,\mu$m and 729 \,nm. The solid vertical line in (a) marks $x = 0.65$. The horizontal dotted line in (b) represent the 100 \,nm thickness of GaAs used in this work. }
\label{fig:CriticalThickness}
\end{figure}

where $I_{\mathrm{PD1}}^{\mathrm{PhC}}$ $\left(I_{\mathrm{PD1}}^{\mathrm{mirror}}\right)$ is the reflected signal intensity of the PhC slab (a mirror of known reflectivity) measured by photodetector 1 (PD1) and $I_{\mathrm{PD2}}^{\mathrm{PhC}}$ $\left(I_{\mathrm{PD2}}^{\mathrm{Mirror}}\right)$ is the reference signal intensity measured simultaneously by photodetector 2 (PD2). This means that we normalize the signal in PD1 by the one in PD2 to account for laser intensity fluctuations. In order to account for any undesired wavelength dependence of the utilized optical components, we independently measure the reflectivity of a mirror of known reflectance, $R_{\mathrm{mirror}}$, in our setup, i.e., ${I_{\mathrm{PD1}}^{\mathrm{Mirror}}}/{I_{\mathrm{PD2}}^{\mathrm{Mirror}}}$ and normalize by this measurement.


\section{Mechanical properties of PhC slabs}

\subsection{Critical thickness, strain and stress}

\fref{fig:CriticalThickness} shows the critical thickness of the two situations we consider in our AlGaAs heterostructures: (a) Al$_{x}$Ga$_{1-x}$As on GaAs, where the AlGaAs layer is used as the sacrificial layer and (b) Al$_{x}$Ga$_{1-x}$As on Al$_{0.65}$Ga$_{0.35}$As, where the Al$_{0.65}$Ga$_{0.35}$As layer is the sacrificial layer assumed to be fully relaxed and the Al$_{x}$Ga$_{1-x}$As layer stands for the GaAs device layer. If the sacrificial Al$_{0.65}$Ga$_{0.35}$As layer in case (b) were not fully relaxed, then the critical thickness were even larger.

As the lattice constant of native GaAs is \unit{5.6533}{\angstrom} ($=a_{\mathrm{film}}$) and of native Al$_{0.65}$Ga$_{0.35}$As is \unit{5.6584}{\angstrom} ($=a_{\mathrm{substrate}}$), a GaAs layer would grow tensile strained on a fully relaxed AlGaAs layer with a strain,  $\epsilon$, of

\begin{equation}
    \epsilon= \frac{a_{\mathrm{substrate}} - a_{\mathrm{film}}}{a_{\mathrm{film}}} = 8.97\cdot10^{-4}.
    \label{eq:Strain}
\end{equation}

\begin{figure}[t!bhp]
\centering\includegraphics{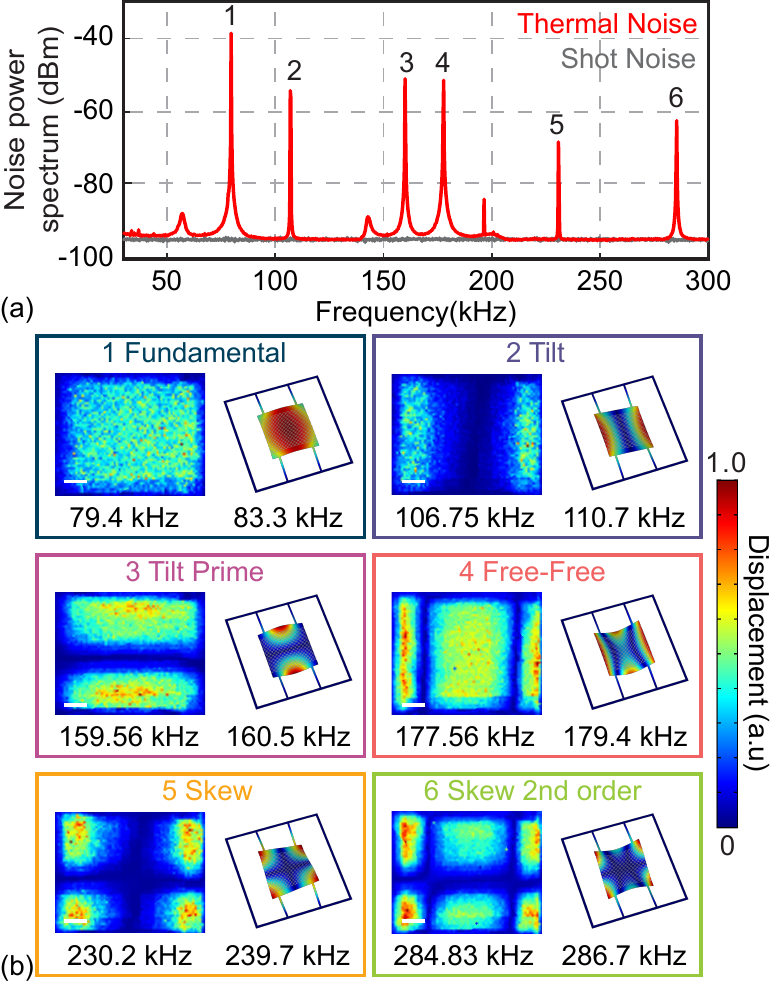}
\caption{Characterization of mechanical modes of a free-free-type PhC slab made from GaAs. (a)~Noise power spectrum of the thermally driven mechanical motion (red) with mechanical modes labeled 1 to 6. (b)~Mechanical-mode tomography of the same device along with FEM-simulated mode shapes and their eigenfrequenices. Scale bar: \unit{10}{\micro\meter}.}
\label{fig:ModeTomo_full}
\end{figure}

This strain leads to a maximal tensile stress $\sigma$ in the GaAs layer  of

\begin{equation}
    \sigma =  E_{\mathrm{GaAs}}\cdot\epsilon = \unit{77.5}{\mega\pascal},
\end{equation}

where $E_{\mathrm{GaAs}}=$ \unit{85.9}{}{\giga\pascal} is the Young's modulus of GaAs. We note that a more detailed calculation of the residual stress in the layer could also consider the strain imprinted in the layer during growth at elevated temperatures and combine these two strains, see Refs.~\cite{Hjort1994,Jeon2000}. As we want to give here only an approximate upper bound on the maximal observable stress in the GaAs layer, this level of detail is not required in our case.

\begin{figure}[t!bhp]
\centering\includegraphics{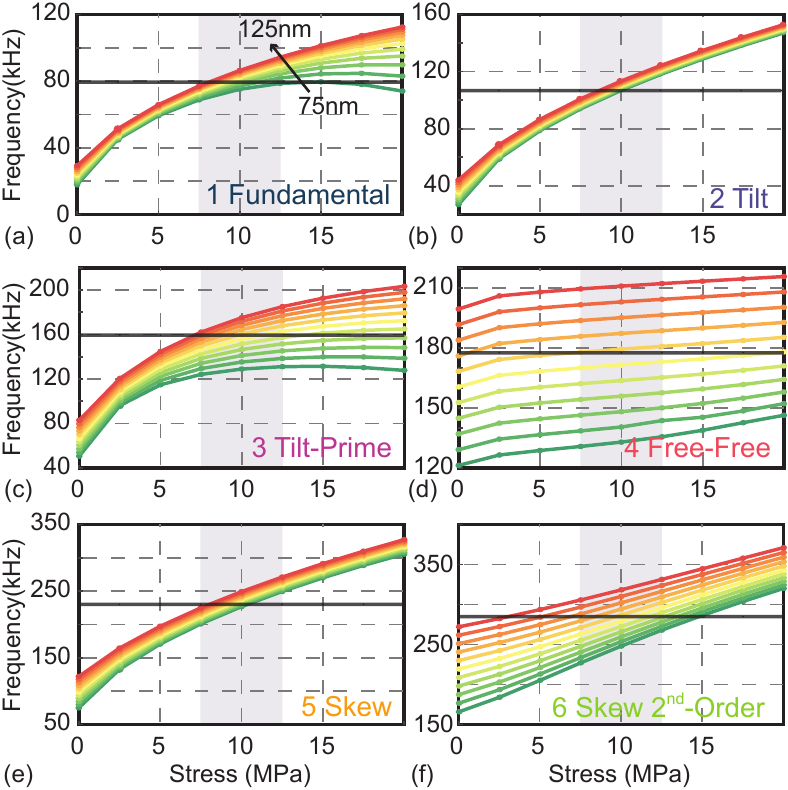}
\caption{FEM simulation results for mechanical frequencies of a $\unit{50\times50}{\micro\meter\squared}$ PhC patterned device for varying stress in the GaAs layer (lines are guides to the eye). The straight horizontal lines show the measured frequencies from Fig.~2(a) of the main text. The thickness varies from \unit{75}{\nano\meter} to \unit{125}{\nano\meter} in steps of \unit{5}{\nano\meter}.}
\label{fig:AllModes_Thickness}
\end{figure}

\subsection{Complete mode tomography of PhC slab}

\fref{fig:ModeTomo_full} shows the noise power spectrum of a free-free-type PhC slab and the mode tomography of the first six mechanical modes. In addition to the modes shown in the main text, we show here also the tilt, tilt-prime and skew 2nd-order mode data and FEM simulation results.

\subsection{FEM-simulations for varying slab thickness and stress}

\fref{fig:AllModes_Thickness} shows FEM simulation results for the mechanical frequencies of different modes when varying the thickness and residual tensile stress in the GaAs slab. We observe that tensile stress leads to an increase of the eigenfrequencies in most cases. However, for the fundamental [\fref{fig:AllModes_Thickness}(a)] and tilt-prime modes [\fref{fig:AllModes_Thickness}(c)] we observe that for thinner slabs the eigenfrequencies reach a maximum at around \unit{15}{\mega\pascal}. 

The fundamental [\fref{fig:AllModes_Thickness}(a)], tilt [\fref{fig:AllModes_Thickness}(b)] and skew modes [\fref{fig:AllModes_Thickness}(e)] show a strong dependence on stress, while they are much less dependent on thickness. From this behavior we can extract a residual tensile stress in the slab of between \unit{7.5}{\mega\pascal} and \unit{12.5}{\mega\pascal} when assuming a realistic slab thickness between \unit{75}{\nano\meter} and \unit{125}{\nano\meter} in order to match to the measured mechanical frequencies.

The opposite behavior is the case for the free-free mode [\fref{fig:AllModes_Thickness}(d)], which is strongly dependent on thickness, but much less on stress. The latter is due to stress relaxation within the pad itself as seen in Fig.~2(c) of the main text and the fact that the free-free oscillation mode is barely affected by the tethers. However, the free-free mode eigenfrequency is directly proportional to the thickness of the PhC slab. Hence, we can use this fact to estimate the slab thickness and get a good match between the simulated and measured frequency at a thickness of \unit{105}{\nano\meter}, when searching in the stress region between \unit{7.5}{\mega\pascal} and \unit{12.5}{\mega\pascal}.  

The tilt-prime [\fref{fig:AllModes_Thickness}(c)] and skew 2nd-order modes [\fref{fig:AllModes_Thickness}(f)] also  depend strongly on thickness and are consistent with the stress and thickness estimates made from the previous four modes.

We, thus, conclude that the slab thickness is around \unit{105}{\nano\meter} and the residual tensile stress in the GaAs layer between \unit{7.5}{\mega\pascal} and \unit{12.5}{\mega\pascal}.

\subsection{Mechanical quality factor data}

\begin{figure}[t!bhp]
\centering\includegraphics{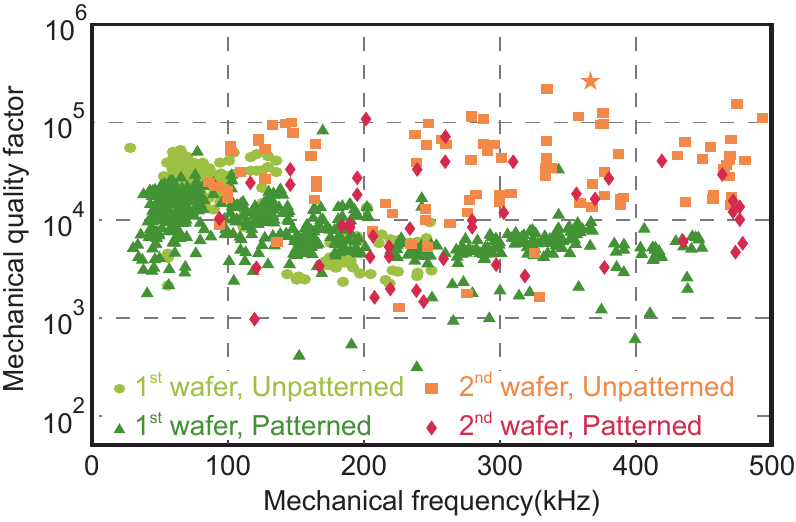}
\caption{Mechanical quality factors for unpatterned and patterned GaAs devices measured at room temperature in high vacuum. We measured 151 devices of different slab length and width. The highest $Q\times f$ product is \unit{9.8\cdot 10^{10}}{\hertz} (star).}
\label{fig:QvsF}
\end{figure}

\fref{fig:QvsF} shows the mechanical quality factors that we determined from ringdown measurements or Lorentzian fits to the NPS measurements for in total 151 single-layer devices fabricated from the first and second wafer, both of patterned and unpatterned GaAs slabs.

\section{Optical properties of PhC slabs}

\subsection{Reflectance map of a PhC slab}

\begin{figure}[t!bhp]
\centering\includegraphics{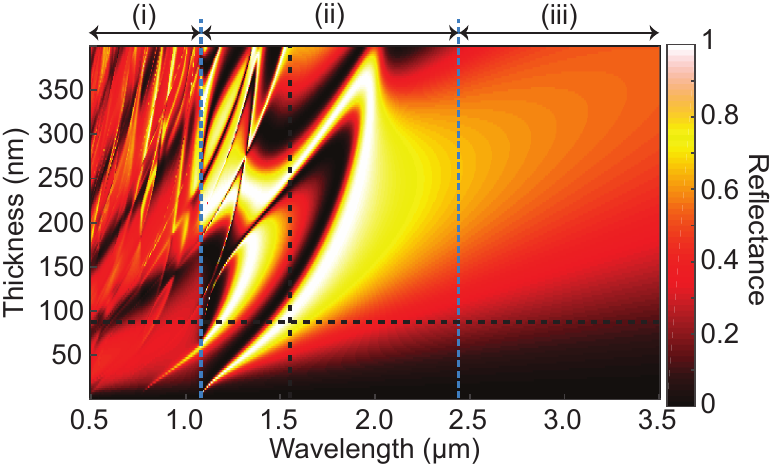}
\caption{Reflectance map of a GaAs PhC slab with $a = \unit{1081}{\nano\meter}$ and $r=\unit{418}{\nano\meter}$ for varying thickness. This map is calculated for plane waves at normal incidence. The black horizontal dashed line represents the slab thickness of \unit{87.5}{\nano\meter}. The black vertical line represents $\lambda = \unit{1550}{\nano\meter}$. The blue dashed lines distinguish between three regimes of optical reflectance: (i) the diffraction regime, (ii) the near-wavelength regime, and (iii) the Fabry-P\'erot regime.}
\label{fig:Thickness_Lambda}
\end{figure}

\fref{fig:Thickness_Lambda} shows the reflectance of a GaAs PhC slab of lattice constant $a=\unit{1081}{\nano\meter}$ and hole radius $r=\unit{418}{\nano\meter}$ for varying slab thickness. We observe three distinct operating regions indicated by the blue dashed lines. In region (i) where $\lambda < a$, diffraction into higher order modes is dominant. In region (ii), where $a < \lambda < a\cdot n_{\mathrm{eff}}$ with $n_{\mathrm{eff}} = (1-\eta)\cdot n_{\mathrm{GaAs}} + \eta\cdot n_{\mathrm{Air}}$ and $\eta = \pi r^2/a^{2}$, the zero-order mode interferes with higher order modes leading to high reflectance regions. In region (iii), where $\lambda > a\cdot n_{\mathrm{eff}}$, a Fabry-P\'erot effect is seen between the two interfaces of the PhC slab \cite{makles_2d_2015}. The black dashed vertical line represents the operating wavelength $\lambda = \unit{1550}{\nano\meter}$ and the black dashed horizontal line represents the slab thickness $h = \unit{87.5}{\nano\meter}$ indicating that we operate in the near-wavelength regime.

\begin{figure}[t!bp]
\centering\includegraphics{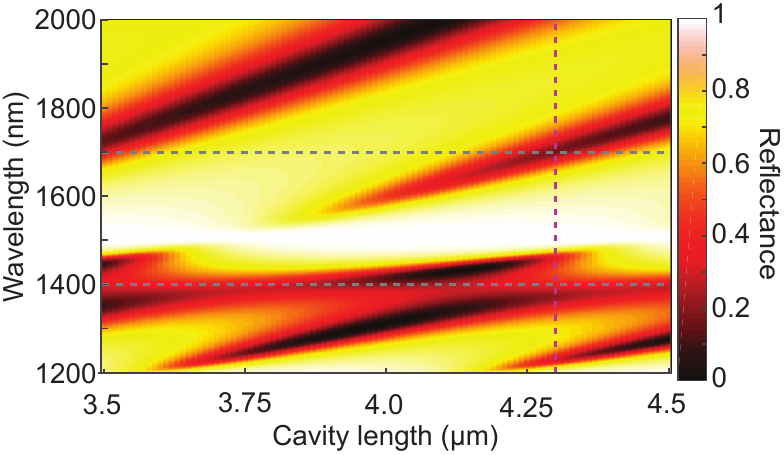}
\caption{Reflectance map of a low-finesse Fabry-P\'erot cavity formed by a PhC slab and GaAs substrate, where the spacing between slab and substrate is changed. This map is calculated for a plane wave at normal incidence. The vertical dashed line represents the spacing used and the horizontal dashed lines represent the wavelength region for simulations shown in the main text.}
\label{fig:slabonsubstAll}
\end{figure}

\subsection{Reflectance map of a PhC slab on a substrate}

In our system, the high-reflectance PhC slab is on top of a GaAs substrate forming a low-finesse Fabry-P\'erot cavity. \fref{fig:slabonsubstAll} shows the reflectance of a PhC slab on top of a GaAs substrate for varying the spacing between slab and substrate. We observe that around $\lambda$ = \unit{1510}{\nano\meter}, i.e., the designed wavelength of the PhC, high reflectance is achieved due to the guided resonance. At wavelengths away from the guided resonance, we observe the Fabry-P\'erot resonances of the low-finesse cavity, visible as tilted black lines.

\begin{figure}[t!bhp]
\centering\includegraphics[width=.8\linewidth]{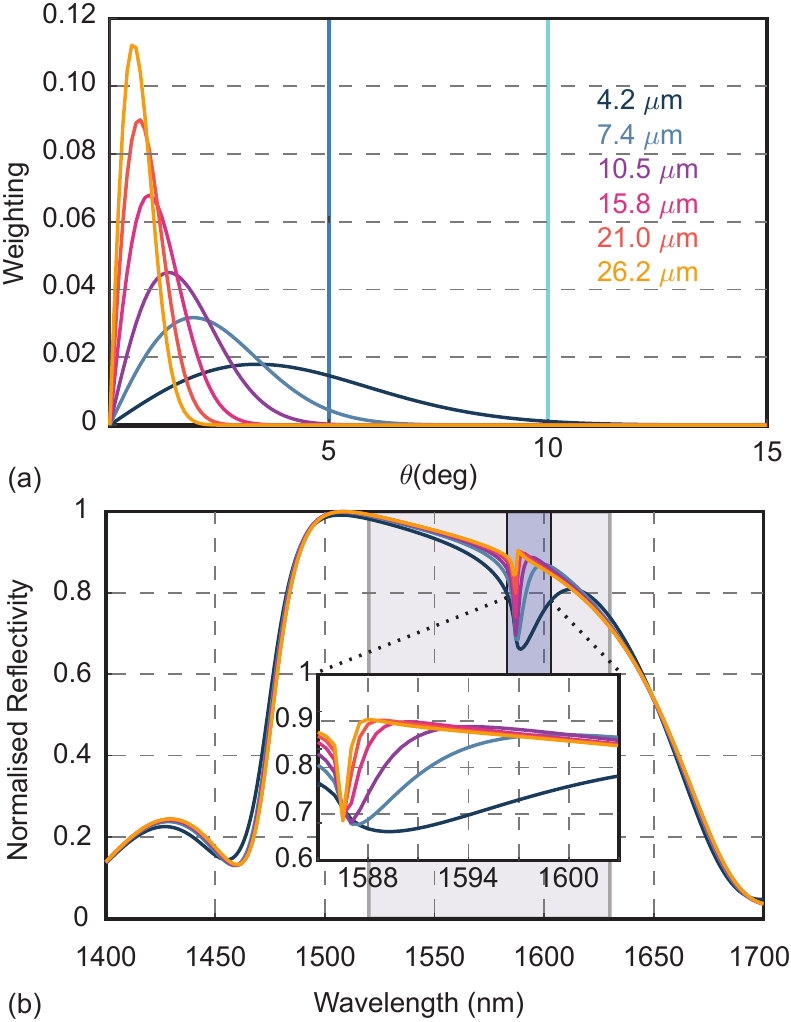}
\caption{Gaussian beam waist dependence on reflectance. (a)~Weighting factor of a plane wave incident at polar angle $\theta$ used for representing Gaussian beams of different waists. (b)~Reflectance spectra of a Gaussian beam of varying waist size incident on an infinite PhC slab on top of a substrate. The inset shows a closer look of the dip.}
\label{fig:Weighting_All_Waists}
\end{figure}

\subsection{Finite beam waist incident on a PhC slab}

Using the angular spectrum representation \cite{novotny2012}, an impinging Gaussian beam can be considered as a sum of plane wave components incident at various angles and weighted by a Gaussian distribution with a standard deviation given by the beam divergence. Rather than using an explicit Gaussian source, we run a series of rigorous coupled wave analysis (RCWA)  simulations, each with a plane wave source, for various angles of incidence and then we construct a weighted superposition of the results. This has the advantage that once we have the simulation results, we need only to change the weights to explore the role of the beam waist.

To see how this works in practice, we note that an arbitrary beam can be split into $s$ (transverse electric) and $p$ (transverse magnetic) components with respect to planes of constant $z$ (we take $z$ as the propagation direction of the beam) \cite{melamed2011}. This division is achieved most succinctly by introducing two unit vectors for a given plane wave: $\hat{n}$, which is normal to the plane of incidence, and $\hat{t}= \hat{k} \times \hat{n}$ defined as 
\begin{eqnarray*}
\hat{n} &= & \frac{1}{k_\parallel}(-k_y \hat{x} + k_x \hat{y}) \\
\hat{t} &= & \hat{k} \times \hat{n} = \frac{k_z}{k_\parallel k} \left(  k_x \hat{x}+ k_y \hat{y} \right) -  \frac{k_\parallel}{k} \hat{z},
\end{eqnarray*}
where $k_\parallel^2=k_x^2+k_y^2$ and the explicit forms are derived by the condition of orthogonality with one another and with the wave vector $\vec{k}$. They can be used, along with the angular spectrum representation, to express the electric field at an arbitrary position along $z$:

\begin{eqnarray*}
\vec{{E}}(\vec{r}_\parallel,z) &= &  \int \frac{d^2k_\parallel}{(2 \pi)^2} \vec{\tilde{E}}(\vec{k}_\parallel)  e^{i\vec{k}\cdot \vec{r}} \\
                               & = & \int \frac{d^2k_\parallel}{(2 \pi)^2} 
\left( \tilde{E}_{s}(\vec{k}_\parallel) \hat{n}(\vec{k}_\parallel)  + \tilde{E}_{p}(\vec{k}_\parallel) \hat{t}(\vec{k}_\parallel)  \right)  e^{i\vec{k}\cdot \vec{r}} \\
                       & = &\vec{E}_{s}(\vec{r}_\parallel,z)   +  \vec{E}_{p}(\vec{r}_\parallel,z),
\end{eqnarray*}
where $\tilde{E}_{s}$ and $\tilde{E}_{p}$ are the projections of the spectral distribution $\vec{\tilde{E}}$ (defined as the spatial Fourier transform of the electric field on the plane $z=0$, we use a tilde to denote Fourier-transformed functions in this appendix) on the unit vectors $\hat{n}$ and $\hat{t}$, respectively.

We assume a material interface that is located at the focus of the beam and in the $xy$-plane. The reflection of the plane wave components of the Gaussian beam is described by the reflection coefficients $r_{s}$ and $r_{p}$, which correspond to polarization perpendicular and parallel to the plane of incidence, respectively. The reflected field at position $z$ can therefore be written as  \cite{novotny2012}
\begin{align}
\vec{{E}}_R(\vec{r}_\parallel,z) & = \\
\nonumber & \int\frac{d^2k_\parallel}{(2 \pi)^2} 
\big( r_{s}\tilde{E}_{s}(\vec{k}_\parallel) \hat{n}(\vec{k}_\parallel)
        +  r_{p}\tilde{E}_{p}(\vec{k}_\parallel) \hat{t}(\vec{k}_\parallel)  \big)  e^{i\vec{k}\cdot \vec{r}}. 
\end{align}

\begin{figure*}[t!bhp]
\centering\includegraphics[width=.82\textwidth]{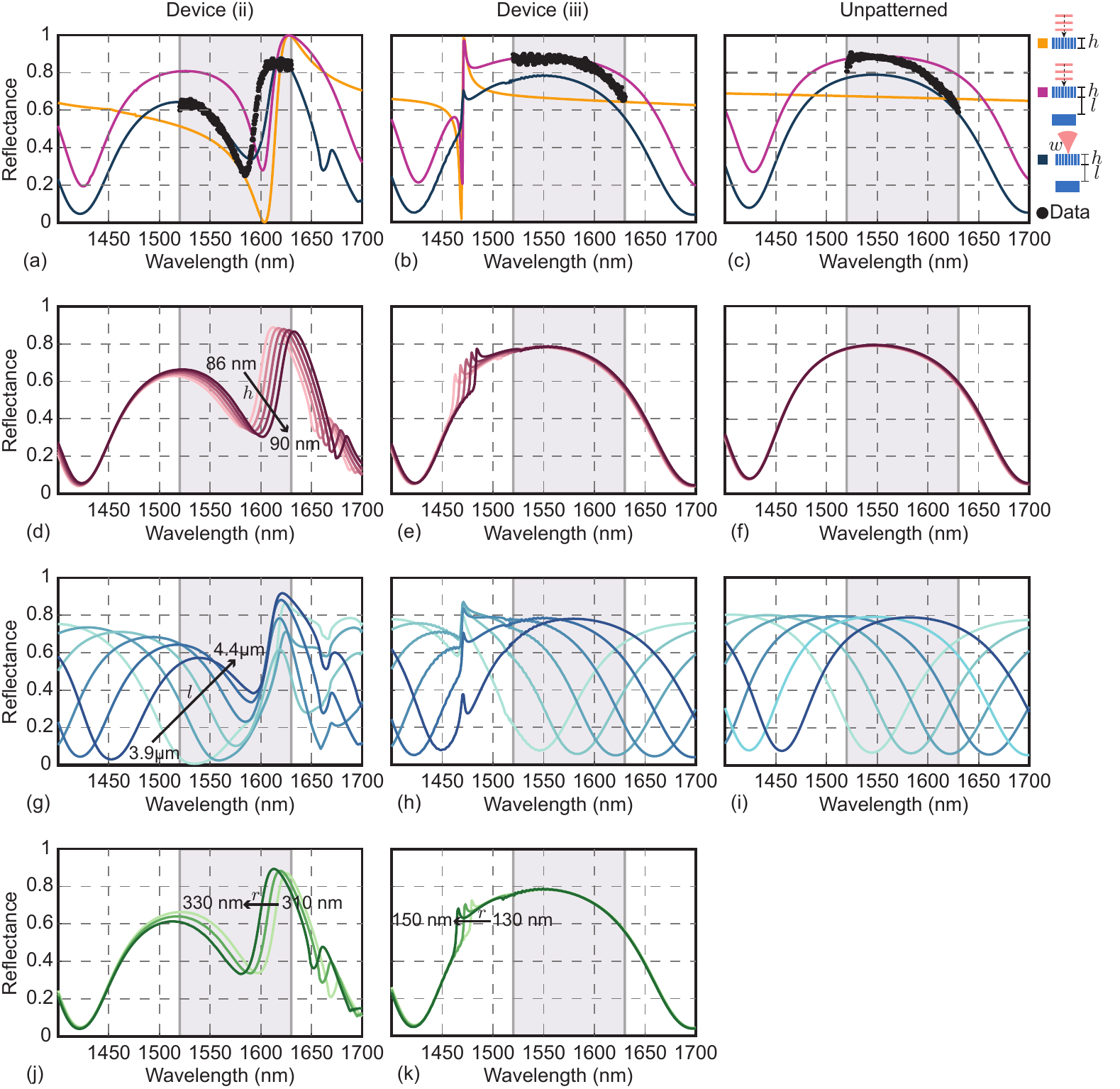}
\caption{(a, b, c) Simulation and measurement of reflectance spectra for devices (ii) and (iii) and an unpatterned device [for the simulations we used $h = \unit{87.5}{\nano\meter}$, $l = \unit{4300}{\nano\meter}$, a beam waist of \unit{4.2}{\micro\meter} and for device (ii) $a = \unit{1452.8}{\nano\meter}$ and for device (iii) $a = \unit{1162.8}{\nano\meter}$]. The measured data (black) are compared to simulated spectra for a plane wave/Gaussian beam of waist \unit{4.2}{\micro\meter} incident on the PhC slab (orange)/(-) or on the slab on top of a GaAs substrate (purple)/(blue). Reflectance spectra for varying (d, e, f) PhC device layer thickness $h$, (g,  h, i) air-gap thickness $l$ and (j, k) PhC hole radius $r$.}
\label{fig:restdevices}
\end{figure*}

From here we take the projection along the propagation direction of the time-averaged Poynting vector and integrate over the transverse direction to get the power flux of the reflected beam. Because of the symmetry of the PhC slab we are concerned with, we may take the beam to be x- or y-polarized for simplicity, e.g. $\vec{\tilde{E}}=\tilde{E} \hat{y}$. After some lengthy calculations and taking the paraxial beam approximation, which is appropriate for the beam waist and wavelength range we consider, we obtain
\begin{equation}
R =  \frac{ \int d\theta d\phi  \
| \tilde{E}(\theta) |^2  \sin(\theta)
\Big(
\sin^2(\phi) |r_{s}|^2 + \cos^2(\phi)|r_{p}|^2
\Big)}{\int d\theta d\phi \
| \tilde{E}(\theta) |^2  \sin(\theta)}, \label{eq:gaussian_weighting}
\end{equation}
which is expressed in terms of the polar angle $\theta$ and azimuthal angle $\phi$. $\tilde{E}(\theta)$ is the electric field distribution of a Gaussian beam at the waist position, which is given by
\begin{equation}
\tilde{E}(\theta) =  \sqrt{2\pi w_0}
 \exp\left[-k^2\sin^2(\theta)  \frac{w_0^2}{4} \right],
\end{equation}
where $w_0$ is the beam waist. Note that we use the opposite spherical coordinate convention to the equivalent expressions given in Ref.~\cite{moura_centimeter-scale_2018}. An illustration of the weighting factor of the reflection coefficients, i.e., $\int |\tilde{E}(\theta)|^2\sin{(\theta)}d \phi$ for varying polar angles $\theta$ is shown in \fref{fig:Weighting_All_Waists}(a).

\fref{fig:Weighting_All_Waists}(b) shows the simulated reflectance spectra of an infinite PhC slab on top of a substrate with a  Gaussian beam of varying waist size impinging on it. We observe the narrowing of the linewidth of the dip with increasing waist size. As we see in \fref{fig:Weighting_All_Waists}(a), as the waist increases the weighting factor for plane waves incident at larger angles curtails their contribution and leads to less dispersion into the guided resonance resulting in a narrowing of the dip.

\subsection{Data and simulations for PhC slabs of different reflectance}

\fref{fig:restdevices} shows the data and simulations for patterned devices (ii) and (iii) and an unpatterned slab. The patterned device (ii) used a PhC pattern with $a = \unit{1452.8}{\nano\meter} $ and $r = \unit{318.8}{\nano\meter}$, aiming at a reflectance of $R=50$\% , and device (iii) used $a = \unit{1162.8}{\nano\meter}$ and $r = \unit{159.18}{\nano\meter}$, aiming at a reflectance of $R=75$\% . 

The deviation of the dip position in \fref{fig:restdevices} (a) is caused by a different thickness assumed in the simulation than the one the actual device has. Recall that the position of the dip depends strongly on the thickness of the PhC, see also \fref{fig:restdevices}(d).

In the reflectance spectrum of an unpatterned device \fref{fig:restdevices}(c,f,i), the Fabry-P\'erot cavity resonance is the dominating feature. We observe that slightly changing the thickness of the slab [\fref{fig:restdevices}(f)] has a minute influence on the spectrum, whereas increasing the air-gap shows the expected spectral shift of the Fabry-P\'erot resonance [\fref{fig:restdevices}(i)]. Interestingly, the reflectance spectrum of the unpatterned device  [\fref{fig:restdevices}(f)] is close to the reflectance spectrum of patterned device (iii) [\fref{fig:restdevices}(e)], apart from the sharp feature in the spectrum occurring for the patterned device. This feature is explained by coupling to a guided resonance of the PhC slab, which is not existent in the unpatterned one. Similar behaviour is seen in the reflectance spectrum of the two devices when varying the air-gap, see \fref{fig:restdevices}(h) and \fref{fig:restdevices}(i).

\clearpage

\bibliography{bib_PhCGaAs}

\end{document}